\documentclass[14pt]{article}

\usepackage{arxiv}

\usepackage[utf8]{inputenc} 
\usepackage[T1]{fontenc}    
\usepackage{hyperref}       
\usepackage{url}            
\usepackage{booktabs}       
\usepackage{amsfonts}       
\usepackage{nicefrac}       
\usepackage{microtype}      
\usepackage{lipsum}
\usepackage{graphicx}
\graphicspath{ {./images/} }
\usepackage{authblk}
\usepackage{amsmath}
\usepackage{amssymb}
\usepackage{setspace}
\title{On the repeatability of turbulence}

\author{Noé Clavier$^1$, Eberhard Bodenschatz$^{1,2,3,*}$ and Florencia Falkinhoff$^{1,*}$}

\begin{document}

\maketitle

\vspace*{-1cm}

$^1${ Max Planck Institute for Dynamics and Self-Organization}, {{Am Faßberg 17}, {Göttingen}, {37077}, {Germany}}  

$^2${ Institute for Dynamics and Complex Systems, Georg August University of G\"{o}ttingen}, {{Friedrich-Hund-Platz 1}, {Göttingen}, {37077}, {Germany}}  

$^3${ Laboratory of Atomic and Solid State Physics, Cornell University}, {{Clark Hall of Science}, {Ithaca}, {14853}, {USA}}  

$^*${ Corresponding authors: \texttt{eberhard.bodenschatz@ds.mpg.de} and \texttt{florencia.falkinhoff@ds.mpg.de} }

\vspace*{0.5cm}

\par\vspace{1ex}

\begin{abstract}
Turbulence has strong and seemingly random fluctuations. Assessing its repeatability is key to predicting flows in technology and nature, much of which decay as viscosity dissipates energy. Much has been done to this end since the work of Lorenz, but mostly in theory and simulations. Here we present experimental results from the Max Planck Variable Density Turbulence Tunnel where we generated decaying turbulence using an active grid, repeating the process with nominally identical initial conditions up to 30,000 times. In contrast with the case of stationary turbulence we found that the energy-carrying large scales show significant repeatability, irrespective of flow development time and turbulence strength. Small scales, however, can effectively be modeled by independent random variables, supporting current numerical approaches in which they are parametrised.

\end{abstract}
\rule{\textwidth}{0.4pt}

\noindent
Turbulent flows have irregular motion and are seemingly unpredictable. They are governed by the Navier-Stokes equations (NSE), which have remained unsolved for over 200 years. Their chaotic behaviour is attributed to non-linear energy transfers across spatial and temporal scales, giving rise to non-Gaussian statistics. Consequently, turbulent velocity fields are commonly treated as random variables since the work by Millionshtchikov and Kolmogorov \cite{millionshtchikov_decay_1939, kolmogorov_local_1941}.  The concept of the spatial energy cascade illustrates the statistical multi-scale nature of turbulence: energy is injected into the system at the energy-containing length scale $L$ and is transferred at a constant energy dissipation rate $\epsilon \propto u^{\prime 3}/L$ (with $u'$ the root mean square of velocity fluctuations) down to smaller scales. This range of scales is the so-called inertial range, and it is here that Kolmogorov's celebrated theory is applied, resulting in an energy spectrum $E(k) \propto \epsilon^{2/3}k^{-5/3}$ (with $k = 2\pi/r$ the wavenumber). In the inertial range, viscous effects are believed to be negligible and the velocity fields to be random. The energy is then dissipated into heat near the Kolmogorov length scale where viscous effects take over. The Taylor-based Reynolds number, Re$_\lambda$, is often used to quantify how strong the turbulence is \cite{Pope2000}. 
While turbulence can have Re$_\lambda$ as low as 100, it can reach 10,000 in the atmosphere.

Although the NSE are deterministic, they are very sensitive to any change in the initial conditions, thus limiting the predictability horizon of turbulent flows (like the weather). Does a small-scale error in the initial conditions contaminate all the scales up to the largest ones, and if so how fast? Is it possible to separate the properties of the large and the small scales? Can the smaller scales be treated as fully random process while only the large scales are simulated? 
Early attempts to answer these questions in the case of homogeneous isotropic stationary turbulence are the seminal works by Lorenz \cite{Lorenz1963, lorenz_predictability_1969} and Leith and Kraichnan \cite{leith_predictability_1972}. They suggested that any initial error cascades up from the small to the large scales without a clear upper limit. Since then, much has been done in more realistic and better resolved simulations \cite{yamazaki_effects_2002, lalescu_synchronization_2013, boffetta_chaos_2017, berera_chaotic_2018, ge_production_2023, bandak_spontaneous_2024, vela-martin_large-scale_2024, vela-martin_predictability_2024, vela-martin_uncertainty_2025}, revealing that in stationary turbulence the error grows exponentially and then linearly until all scales are affected after a few large-scale eddy turnover times $\tau \equiv L/u'$. However, many flows depart from this ideal situation in ways that might affect qualitatively their predictability: geometry \cite{lo_predictability_2015}, stratification \cite{diaz_predictability_2024}, or simply turbulence decay. In many flows, indeed, no new energy is supplied at $t>0$ and turbulence decays with time as the energy is dissipated, starting from the smallest scales. Such decaying flows include wind turbine wakes, wind tunnel flows and certain geophysical flows. Ref. \cite{metais_statistical_1986} found through numerical simulations that the error grows much slower in that case, but couldn't extend their observations until the largest scales are reached. Furthermore, much less is known in experiments where initial conditions are hard to reproduce, as well as about the largest scales of the flow which might be more repeatable (see \cite{borodulin_experimental_2011, borodulin_experimental_2013} for the case of boundary layer instability) and enable to forecast certain types of extreme events \cite{vela-martin_large-scale_2024}.

Here we report on an experimental study of the repeatability of decaying turbulent flows at all scales.
In this work, we carried out wind tunnel experiments where an active grid is used to generate well-controlled decaying turbulence and as accurately as possible repeat the same flow conditions up to $N = 30,\!000$ times, in a manner analogous to Monte Carlo ensembles in simulations \cite{leith_theoretical_1974, vela-martin_predictability_2024, vela-martin_uncertainty_2025}. Analysis of the subsequent ensemble statistics of turbulent fluctuations is presented and, in particular, we show that the ensemble average of the $N$ velocity fluctuations time-series does not vanish but instead retains most of the large scale structures, irrespective of the flow development time and strength of turbulence (Re$_\lambda$). These observations imply  that the large scales of the flow can be reliably repeated to a significant degree.
On the other hand, scales in most of the inertial range of turbulence are filtered out by the process of ensemble averaging. The data together with a model of the effect of ensemble-averaging directly confirm that the smaller scales of turbulence (i.e., most of the inertial and dissipation ranges) can effectively be modelled by independent random variables.


\subsection*{Large ensembles of wind tunnel experiments}

Experiments were carried out in the Max Planck Variable Density Turbulence Tunnel (VDTT), a closed-loop wind tunnel that can be pressurised up to 19 bar and filled with various gases (see \cite{bodenschatz_variable_2014} and the Methods section). Using an active grid (grid in the following) to control inflow conditions \cite{griffin_control_2019}, we repeatedly generated decaying turbulent flows with the same initial conditions. More specifically, the protocol of duration $T$ which controlled the grid was repeated $N$ times while the flow conditions (velocity, viscosity, geometry) were left unchanged, except for small-scale eddies which appeared in the tunnel before they pass through the grid and added some error to the initial conditions. $N$ was taken up to $30,\!000$ times and $T$ was in the range $[3.9, 89]\tau$ (eddy turnover times) and $[6, 200]$ seconds. Henceforth, each $T$-long repetition of this procedure is referred to as a \textit{realisation}, and we call \textit{experiment} the ensemble of $N$ realisations with the same grid protocol. We denote time averages by $\overline{\,\cdot\,}$ and ensemble averages over $N$ realisations $\langle \cdot \rangle_N$. For each realisation $n$, we measured the flow's streamwise velocity $v_n(t)$ at a distance $x$ downstream from the grid, corresponding to a development time $t_\text{dev}=x/U$ where $U = \langle\overline{v_n}\rangle_N$ is the mean flow speed. We denote $u_n(t) \equiv v_n(t) - U$ the velocity fluctuations for each realisation. The ensemble-averaged velocity fluctuations are then defined by $\langle u\rangle_N(t) \equiv N^{-1}\sum_{n=1}^N u_n(t)$, and flow characteristics for one experiment ($u'$, Re$_\lambda$, $E(f)$) are averaged over all its realisations (e.g. $u' = \langle(\overline{u_n^2})^{1/2}\rangle_N$).
In most cases, the $N$ realisations were performed one after the other, that is, the grid forcing was periodic. The protocol duration $T\gg\tau, t_\mathrm{dev}$ ensured sufficient statistical significance of $u_n$ and that the flow was locally indistinguishable from an aperiodic flow. In addition, experiments where the $N$ realisations were separated by fixed or even random delays showed no difference to the periodic ones. In this work, we analyse a total of 36 experiments with turbulence intensity Re$_\lambda$ ranging from 150 to 3100 with different isotropic and anisotropic protocols.

\begin{figure}
\centering
\includegraphics[width=0.5\textwidth]{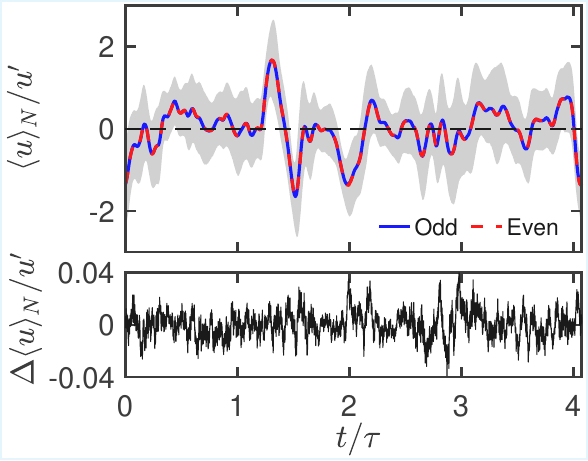}
\caption{\textbf{Ensemble-averaged velocity fluctuations.} Top: Ensemble-average velocity fluctuations of the $N/2=14,\!999$ odd and even realisations (in solid blue line and dashed red line, respectively), normalised by $u'$. The shaded area represents a $\pm1$ standard deviation $\sigma(t) \equiv \langle (u_n-\langle u\rangle_N)^2\rangle_N^{1/2}$. Bottom: Difference between the odd- and even-realisations ensemble-averaged signals, $\Delta\langle u\rangle_N \equiv \langle u\rangle_{N/2,\text{odd}} - \langle u \rangle_{N/2,\text{even}}$.}
\label{fig:phaseAverage}
\end{figure}

The ensemble-averaged velocity fluctuations for one experiment with $N=29,\!998$ are shown in Figure \ref{fig:phaseAverage}. Odd and even realisations are displayed separately for comparison. If none of the flow features were repeatable, the ensemble-averaged signal would tend to zero as $N \rightarrow \infty$. On the contrary, the signal converges to non-zero values with fluctuations of  order $u' = 0.21$~m\,s$^{-1}$, in line with results reported in \cite{borodulin_experimental_2011} for boundary layer instability. This implies that some part of the flow is—to some extent—reproducible. The velocity at any given time $t$ fluctuates significantly from one realisation to the next (following Gaussian or skewed distributions, not shown), as indicated by the large 1-standard-deviation area in Fig. \ref{fig:phaseAverage}. Nevertheless, some probabilistic forecasts can be made. For example, 97\% of realisations of this experiment have positive fluctuations at $t/\tau = 1.3$, with probability $p(u>u^\prime | t/\tau=1.3) = 75\%$, whereas $p(u>u^\prime) = 15\%$ for all other times.


\subsection*{Repeatability across turbulent scales}

\begin{figure}[t]
\centering
\includegraphics[width=\textwidth]{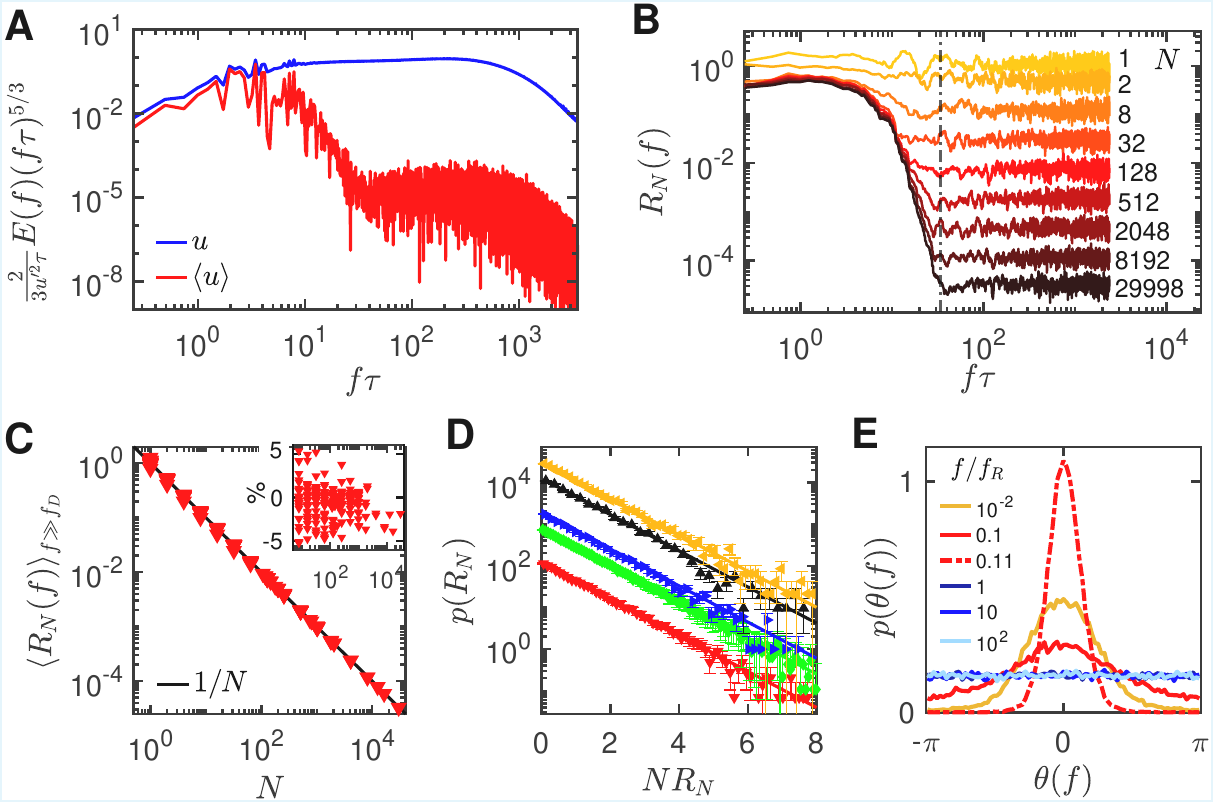}
\caption{
\textbf{Energy spectra and comparison to a model of purely non-repeatable flow.}
\textbf{(A)} Compensated energy spectrum of the velocity fluctuations $u$ (in blue), and of the ensemble-averaged velocity fluctuations $\langle u\rangle_{N}$ (in red), as a function of the normalised frequency $f\tau$. $N=29,\!998$.
\textbf{(B)} Energy attenuation factor $R_{N}(f)$ for this experiment for various numbers of realisations $N$ (indicated in plot) over which the ensemble-average is computed. The vertical line indicates $f_R\tau$. Data is filtered for visualisation purpose only.
\textbf{(C)} Mean energy attenuation factor $\langle R_{N}\rangle$ in the range where it fluctuates around a plateau for all 32 experiments at a fixed $x$, compared to the model's prediction $\langle R_{N}\rangle=1/N$ (solid line). Inset: relative deviation of the experimental data to the model in $\%$ as a function of $N$.
\textbf{(D)} Probability density distributions of $R_N$ in the plateau range for five experiments with different Re$_\lambda$ and grid protocols (from bottom to top, $N = 118,750,1\,798,12\,500,29\,998$). Solid lines: model $p(R_N) = N\exp(-NR_N)$.
\textbf{(E)} Probability distribution of the phases of a few Fourier modes (indicated in plots in units $f_R$) in the same experiment as (A), centred around their circular mean. For $f/f_R \ll 1$, the distributions are close to wrapped normal distributions (orange, red). As $f$ increases, the distributions flatten and become indistinguishable from uniform distributions when $f/f_R = 1$ (blue curves).
}
\label{fig:spectrum}
\end{figure}

In order to assess the repeatability of each flow scale, we calculate the energy spectrum $E$ of the velocity fluctuations as a function of frequency $f$ for each experiment (averaged over all $N$ realisations), as well as the spectrum $E_{\langle u\rangle_N}(f)$ of the ensemble-averaged velocity fluctuations $\langle u\rangle_N$. As a consequence of Taylor's frozen flow hypothesis, small frequencies correspond to the large spatial scales, and vice-versa. This transformation is however not necessary to the present analysis.
Figure \ref{fig:spectrum}A shows such spectra compensated by the expected Kolmogorov scaling for one experiment (Re$_\lambda = 390, N=29,\!998$). The compensated $E(f)$ displays a plateau characteristic of the inertial range of fully developed turbulence, here over almost two decades. On the other hand, the spectrum of the ensemble-average $\langle u\rangle_N$ presents three striking features: (i) the slow, large scales are only weakly attenuated by ensemble-averaging (ii) the faster, smaller scales are suppressed, starting roughly at the beginning of the inertial range, and (iii) the spectrum in this range is very ``noisy". Ensemble-averaging effectively acts as a low-pass filter, consistent with findings in simulations \cite{vela-martin_predictability_2024}, but one that introduces noise in the spectrum. To better quantify these features, we introduce the attenuation factor $R_N(f) \equiv E_{\langle u\rangle_N}(f)/E(f)$ which measures how much energy is left in the ensemble-averaged signal.

Fig. \ref{fig:spectrum}B shows $R_N(f)$ for different $N$. As $N\rightarrow \infty$, the small-scale (large $f$) attenuation plateaus at a decreasing value, whereas the large-scale (small $f$) attenuation saturates at $\approx 0.4$. The transition between the two, at $f\tau \sim {O}(10)$ (dashed vertical line), follows a power law with a negative exponent decreasing with $N$. The intersection of this power law with the asymptotic plateau at $f \equiv f_R$ defines a time scale $f_R^{-1}$ (equivalently a length scale $L_R = U/2\pi f_R$) which we estimate in the large $N$ limit (see the Extended Data). $R(f\ll f_R) = O(1)$ implies that below $f_R$, the flow is repeatable to an extent. On the other hand, above $f_R$, the flow can be described by independent random variables and is not repeatable. Indeed, we show that the asymptotic behavior of $R_N(f)$ can be explained by a simple statistical model, relying on two assumptions only: (i) the energy spectra of all realisations are strictly identical, i.e. $\forall n, E_{u_n}(f) = E(f)$, and (ii) the phases $\theta_n(f)$ of the Fourier mode $f$ of each realisation $n$ are independent, identically distributed random variables following a uniform distribution for $f > f_R$. Its main result is that the distribution of $R_N(f)$ is independent of $f$ and exponential for $N\gg 1$ (see Supplementary Information):
\begin{equation}
    \label{eq:R_N}
    \forall f> f_R,~~ \mathbb{E}[R_N(f)] = \frac{1}{N},~p(R_N(f)) = N\exp(-NR_N(f)).
\end{equation}
Figure \ref{fig:spectrum}C shows that this prediction of $\langle R_N\rangle$ matches the average value of $R_N(f)$ for $f>f_R$ for all experiments, over more than 4 decades in $N$, with a relative error within $\pm5\%$ (shown in the inset). It must be pointed out that the accuracy of the measurements of $R_N$ for $N \gtrsim 10^4$ may be less, as  $E_{\langle u\rangle_N}(f)$ is then very small in absolute value. Additionally, Fig. \ref{fig:spectrum}D shows that the exponential distributions of $R_N$ in its plateau range agree with equation \ref{eq:R_N}. Nonetheless, this evidence is not entirely sufficient to demonstrate assumption (ii), for peculiar multi-peaked distributions of the $\theta_n(f)$s could result in the same behaviour of $R_N$. This case can be discarded by a closer examination of these distributions, shown in Fig. \ref{fig:spectrum}E. For $f>f_R$, they are indistinguishable from uniform. This result is a direct, experimental confirmation that turbulence can effectively be modeled by independent random variables for all $f>f_R$, that is, at the smaller spatial scales.

Conversely, at the lower frequencies $f \ll f_R$ (large spatial scales) the distribution of $\theta_n(f)$ is peaked, indicating that all realisations of the flow have similar phases, i.e. these scales are repeatable to a large extent. With increasing $f$, the distribution overall flattens until it becomes indistinguishable from uniform. However, this is not monotonic, as Fig. \ref{fig:spectrum}E shows: consecutive Fourier modes can display very different peak width. Since they are two ways to quantify the same phenomenon, the $f$'s for which the $\theta_n(f)$-distribution is particularly wide (e.g. $f/f_R = 1.1$ in Fig. \ref{fig:spectrum}E) naturally correspond to modes where $E_{\langle u\rangle_N}(f)$ is significantly less than $E(f)$ (Fig. \ref{fig:spectrum}A). Although smaller then, $R_N(f)$ remains $O(1)$. This indicates that some specific large scales of our flow contain random and repeatable components with comparable energies, whereas other large scales are much more repeatable. What specifically dictates these scales deserves further investigation.

\begin{figure}
\centering
\includegraphics[width=\textwidth]{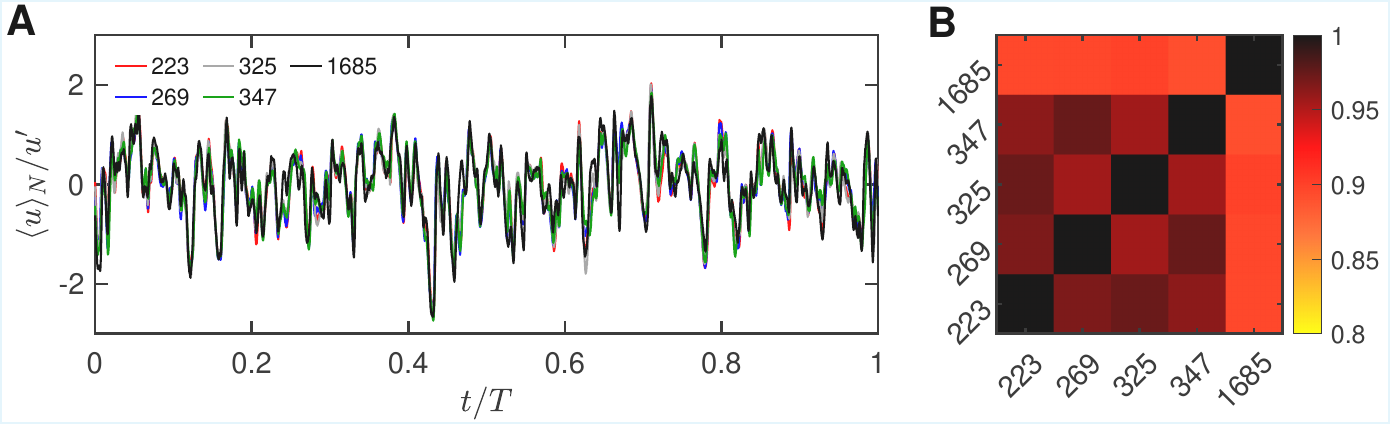}
\caption{
\textbf{Independence of the rescaled ensemble-averaged fluctuation time series and of the Reynolds number.}
\textbf{(A)} Ensemble-averaged velocity as a function of time for five different experiments in rescaled time units, for different Re$_\lambda$ (indicated in the figure). $T$ ranges from 25 to 50 seconds, $u'$ from 0.14 to 0.29 m\,s$^{-1}$. For the four lowest Reynolds numbers only $U$ was changed, with $f_\text{grid}/U = \text{cst}$. Re$_\lambda = 325$ and $1685$ differ only by the viscosity $\nu$.
\textbf{(B)} Pearson correlation coefficient between each pair of ensemble-averaged signals.
}\label{fig:ReInvariance}
\end{figure}

Furthermore, these observations are independent of the strength of turbulence. We corroborate this by changing either the grid frequency $f_\text{grid}$ while keeping the ratio $f_\text{grid}/U$ constant, or the viscosity $\nu$. In both cases Re$_\lambda$ is modified but the characteristic length scales of the flow are not, nor are the time scales when rescaled by $T$ or $f_\text{grid}^{-1}$|the flow is self-similar. In Figure \ref{fig:ReInvariance}A, we plot $\langle u\rangle_N/u'$ as a function of $t/T$. The signals collapse remarkably well in spite of variations of the strength of turbulence (Re$_\lambda$) by a factor of 8. As Fig. \ref{fig:ReInvariance}B shows, the Pearson correlation coefficients are greater than $0.89$ ($0.95$ for the lower Re$_\lambda$s). In other words, in rescaled units the repeatable components of the flow are independent of Re$_\lambda$.

These findings can be related to the modelling carried out in Large Eddy Simulations (LES) and Reynolds-Averaged Navier-Stokes simulations (RANS). In LES, the velocity is decomposed into a resolved large-scale component, and a sub-grid scale residual component acting on the large scales only via its statistical properties. In our experiments, the ensemble-averaged fluctuations shown in Fig. \ref{fig:ReInvariance}A can be interpreted as the resolved component normally used in LES \cite{vela-martin_predictability_2024}. On the other hand, the filtered-out part $u_n(t)-\langle u\rangle_N(t)$ corresponds to the residual component which needs modelling. If one is only interested in the flow components that are repeatable, the proposed $L_R$ would be a natural choice for LES cut-off length scale—or frequency. Moreover, when the grid resolution is increased in LES (and so is Re$_\lambda$), the resolved signal is not expected to change. Fig. \ref{fig:ReInvariance} can thus be interpreted as an experimental evidence of why this can be expected to work: the repeatable part of the velocity field remains Reynolds-independent under the right scale-invariant transformation. Likewise, in RANS turbulent fluctuations are modeled through time averaging, not unlike what we report in these experiments.

\subsection*{Time evolution in decaying turbulence}

Predictability of turbulence can only be assessed by looking at the time evolution of the flow. Do the repeatable features vanish as the flow development time $t_\text{dev}$ increases? To answer this question, we first need to quantify the amount of unpredictable features in the total flow. We do so through the ratio of the turbulent kinetic energy (TKE) of the random, non-repeatable components ($e_\text{rand}$) to the total total turbulent kinetic energy ($e_\text{tot}$), which reads:
\begin{equation}
    \label{eq:erand}
    \frac{e_\text{rand}}{e_\text{tot}} = \lim_{N\to\infty} \left[1 - \frac{\overline{\langle u\rangle_N^2}}{u'^2} \right].
\end{equation}
$e_\text{rand}$ is actually equivalent to the more common total error energy (see the Extended Data). The limit effectively converges for $N \gtrsim 10^2$, and our experiments show that $e_\text{rand}/e_\text{tot}$ ranges from 90\% to as little as 35\%, depending on the initial conditions (see Fig. \ref{fig:ErandEtot_convergence} from the Extended Data and Table \ref{tab:list_exps} from the Supplementary Information).

Numerous studies of stationary turbulence suggest that $e_\text{rand}/e_\text{tot}$ should approach 1 as time increases. More specifically, after some transient time the error in the initial conditions—which was initially confined to small scales—contaminates at time $t_\text{dev}$ all scales up to $L_R \propto \epsilon^{1/2}t_\text{dev}^{3/2}$; and as a consequence the random energy ratio increases linearly with time \cite{Lorenz1963, boffetta_chaos_2017, berera_chaotic_2018, ge_production_2023}
\begin{equation}
    \frac{e_\text{rand}}{e_\text{tot}} = G e_\mathrm{tot}^{-1}\epsilon t_\text{dev},
    \label{eq:ErandEtot_scaling}
\end{equation}
where $G$ is a dimensionless constant a priori depending on the initial conditions and the Reynolds number. In the present experiments however, turbulence it not stationary but decays. After a short transient regime, the flow is characterised by a self-similar decay of the turbulent kinetic energy $e_\text{tot} \propto t_\text{dev}^{-n}$ and simultaneously of the dissipation rate $\epsilon \propto t_\text{dev}^{-n-1}$, resulting in a growing eddy turnover time $\tau \propto t_\text{dev}$ \cite{Pope2000, sinhuber_decay_2015}. Since it is self-similar, the flow does not possess any new time or length scale compared to the stationary case and, consequently, we expect the afore-mentioned scalings to still hold. However, Eq. \ref{eq:ErandEtot_scaling} now implies that $e_\text{rand}/e_\text{tot}$ remains constant with time, ensuring a non-vanishing degree of repeatability of the flow features as long as the turbulence lives.

\begin{figure}[t]
\begin{flushleft}
    \large{\textsf{\textbf{A~~~~~~~~~~~~~~~~~~~~~~~~~~~~~~~~~~~~~~~~~~~~B}}}
\end{flushleft}
\centering
\vspace*{-0.5cm}
\includegraphics[width=0.43\textwidth]{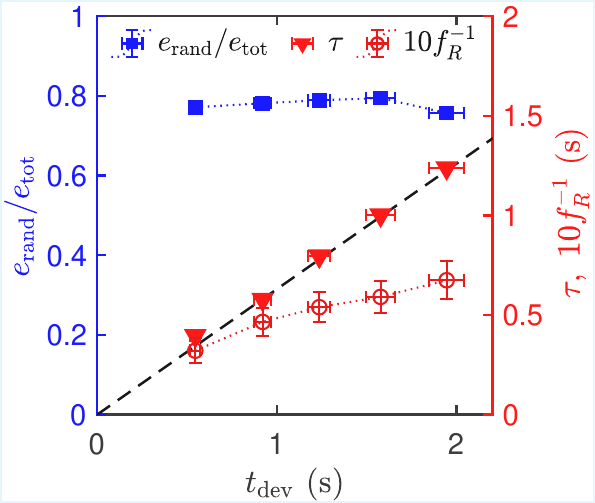}
\hfill
\includegraphics[width=0.56\textwidth]{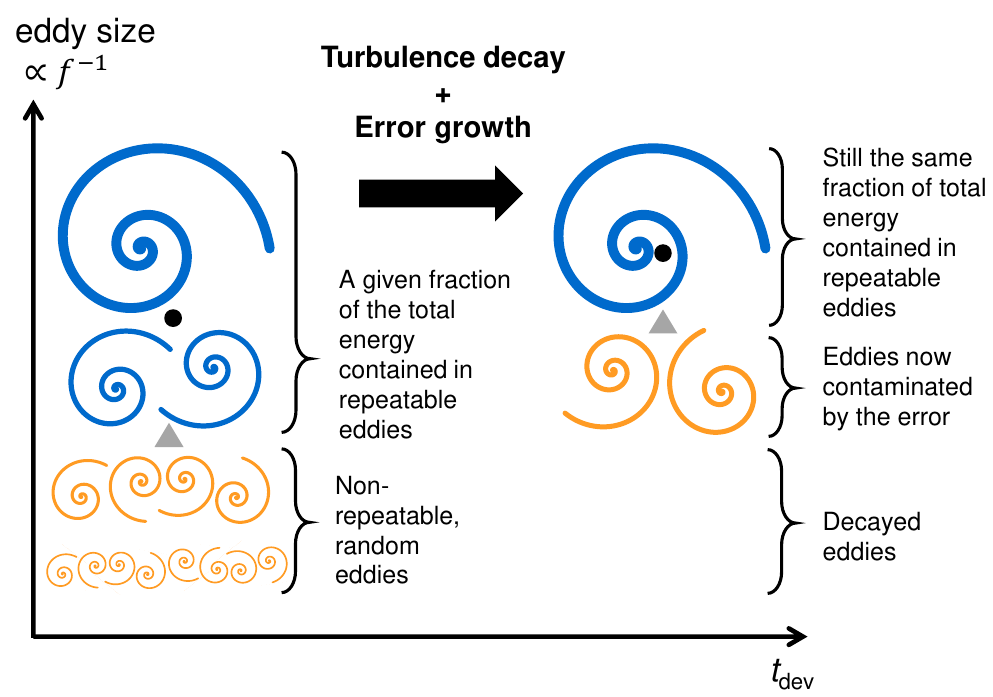}
\caption{
\textbf{Evolution of repeatability with time in decaying turbulence.}
\textbf{(A)} Evolution of the fraction of random energy in the flow (left axis), and of the integral time scale $\tau = u'^3/\epsilon$ and the repeatability time scale $f_R^{-1}$ (right axis), as turbulence decays with development time. Error bars on $t_\text{dev}$ are deduced from the uncertainty on the measure of $U$, and those on $f_R$ from the fits to estimate $f_R$ (see Fig. \ref{fig:f_R}). The dashed-line shows $0.63 t_\text{dev}$.
\textbf{(B)} Schematic describing the simultaneous growth of the energy-containing scales and of the scales contaminated by uncertainty as turbulence decays. The grey triangles and the black circles indicate eddy size corresponding to $f_R^{-1}$ and $\tau$, respectively.}
\label{fig:time_evolution}
\end{figure}

We verified this assertion in a set of experiments where we varied $t_\text{dev}$ by a factor 3 by measuring the velocity at different downstream positions. Figure \ref{fig:time_evolution}A shows close agreement of these experiments with the proposed scaling: the eddy turnover time $\tau$ increases linearly with time, the characteristic error scale $f_R^{-1}$ increases but slower, and the random energy ratio is constant. In decaying turbulence, the error cascade is hence qualitatively different from the stationary case. Likewise, we expect the error to grow exponentially fast at short times (shorter than those studied here), as quantified by the Lyapunov exponents \cite{boffetta_chaos_2017, berera_chaotic_2018, ma_effect_2024}.
However, as the error cascades up throughout the inertial range, the growth in time and length scales of the decaying turbulence competes with the growth of the error. The error contaminates more and more scales, which however are depleted of their energy by dissipation in the meantime. This concept is illustrated schematically in Fig. \ref{fig:time_evolution}B. We find that the two phenomena compensate, in such a way that the remaining TKE seems to always keep a memory of the initial conditions.

A question that remains open is what defines the prefactor $G$ in equation \ref{eq:ErandEtot_scaling}. We found values ranging from 0.1 to 1, of similar order of magnitude as previously reported values \cite{leith_predictability_1972, boffetta_chaos_2017}, but we are not able to relate them to inflow conditions in the present study. Elucidating this question would be necessary to quantify a priori the predictability of a given decaying flow.


\subsection*{Discussion and outlook}

In conclusion, these experiments shed new light on the predictability of freely-evolving, decaying turbulent flows. Using the active grid to reproduce nearly the same initial conditions repeatedly, we obtained ensemble statistics with an unprecedented sample size in experiments. Our data shows that large, energy-containing scales of the flows are repeatable to an extent, regardless of Reynolds number and flow development time, a feature clearly distinct of the case of stationary turbulence. 

Therefore, if ensemble statistics are available, probabilistic forecasts on later realisations of the same flow can be made, which supports the use of ensemble-based weather forecasts \cite{pantillon_2017, palmer_2017, yousefnia2025}. Besides, wind turbine wakes are an example of decaying turbulence, which in addition stems from an almost periodic excitation (the propeller's rotation). It seems therefore plausible that the specific large-scale eddies experienced by the downstream wind turbines, known to limit dramatically their performance, are predictable individually to a large extent. These remarks also apply for aerodynamics testing in wind tunnels in which models of wind turbines or civil engineering structures experience decaying turbulence \cite{howard_characterizing_2015, grunwald_effect_2026, kildal_use_2023}. A number of geophysical flows present such characteristics too, but many are stratified, 2-dimensional and/or rotating, and the situation there might yet differ. 

Our experiments nevertheless confirm directly that the smaller scales (most of the inertial range and the dissipation range) can be modeled by independent random variables, as proposed by Kolmogorov's phenomenological theory. Further research would be needed to identify the origin of the error in the initial conditions, here likely due to remaining eddies before the grid, but which could be in other cases be related to thermal noise \cite{bandak_spontaneous_2024, Liao_2025} or other kinds of small scale perturbations. As a consequence, ensemble-averaging effectively acts as a low-pass filter which partially preserves the ``smoother" part of the velocity fluctuations but completely eliminates the ``rougher" ones. The filtered signal can be interpreted as the resolved component in LES, and the proposed cut-off frequency $f_R$ between the random and the (partially) repeatable components could prove useful to estimate the optimal grid scale in LES. In particular for weather and climate models, if high-resolutions simulations are to be achieved \cite{Palmer2019}, they might not need to model the turbulent stochastic processes at small scales or other kinds of extreme phenomena \cite{Stevens2013}. The same might be said for other systems that involve turbulent flows, such as cloud modelling, where the scales range from micrometers to meters \cite{Bony2015}.

However, what determines the constant amount of repeatable and possibly predictable features in a flow (that is, $e_\text{rand}/e_\text{tot}$) remains an open question. This invites new perspectives on how we think about the apparently random nature of turbulent flows, which has coexisted with the deterministic Navier-Stokes equations for more than two centuries.



\vspace{1.5cm}

\bmhead{Supplementary information}
Additional information about the methods and additional figures can be found in the Supplementary Information.

\bmhead{Acknowledgements}
The VDTT and its electronics were built, operated or maintained by A. Renner, A. Kopp, A. Kubitzek, H. Nobach and the machinists at the MPI-DS. F. Lachauss\'{e}e and G. Bewley have been the firsts to try using periodic protocols of the active grid in the VDTT. We are grateful to M. Vergassola, J. Bec, L. Szekelyhidi Jr., G. Voth, C. Brunner, M. Grunnwald, H. Kim, A. Pumir, P. Spichtinger and F. Hoffmann for discussions.

\section*{Declarations}

\begin{itemize}
\item \textbf{Competing interests:} The authors declare no competing interest.
\item \textbf{Inclusion and Ethics:} The published results are purely technical and did not involve any human, animal or biological samples.  All authors of this study have fulfilled the criteria for authorship, and responsibilities were agreed among the collaborators. 
\item \textbf{Author contribution:} N.C. carried out the experiments, analysed the data and developed the theoretical model. F.F. and E.B. designed the project. F.F. supervised the project. N.C., F.F. and E.B. discussed and interpreted the results, wrote and reviewed the article.
\item \textbf{Data availability:} The datasets  reported in this publication are available from the corresponding authors upon reasonable request.
\item \textbf{Code availability:} The post-processing codes used for this publication are available from the corresponding authors upon reasonable request.
\end{itemize}

\bibliographystyle{ieeetr}
\bibliography{bibliography}

\backmatter

\subsection*{Methods}

\begin{figure}[t] 
	\centering
	\includegraphics[width=0.8\textwidth]{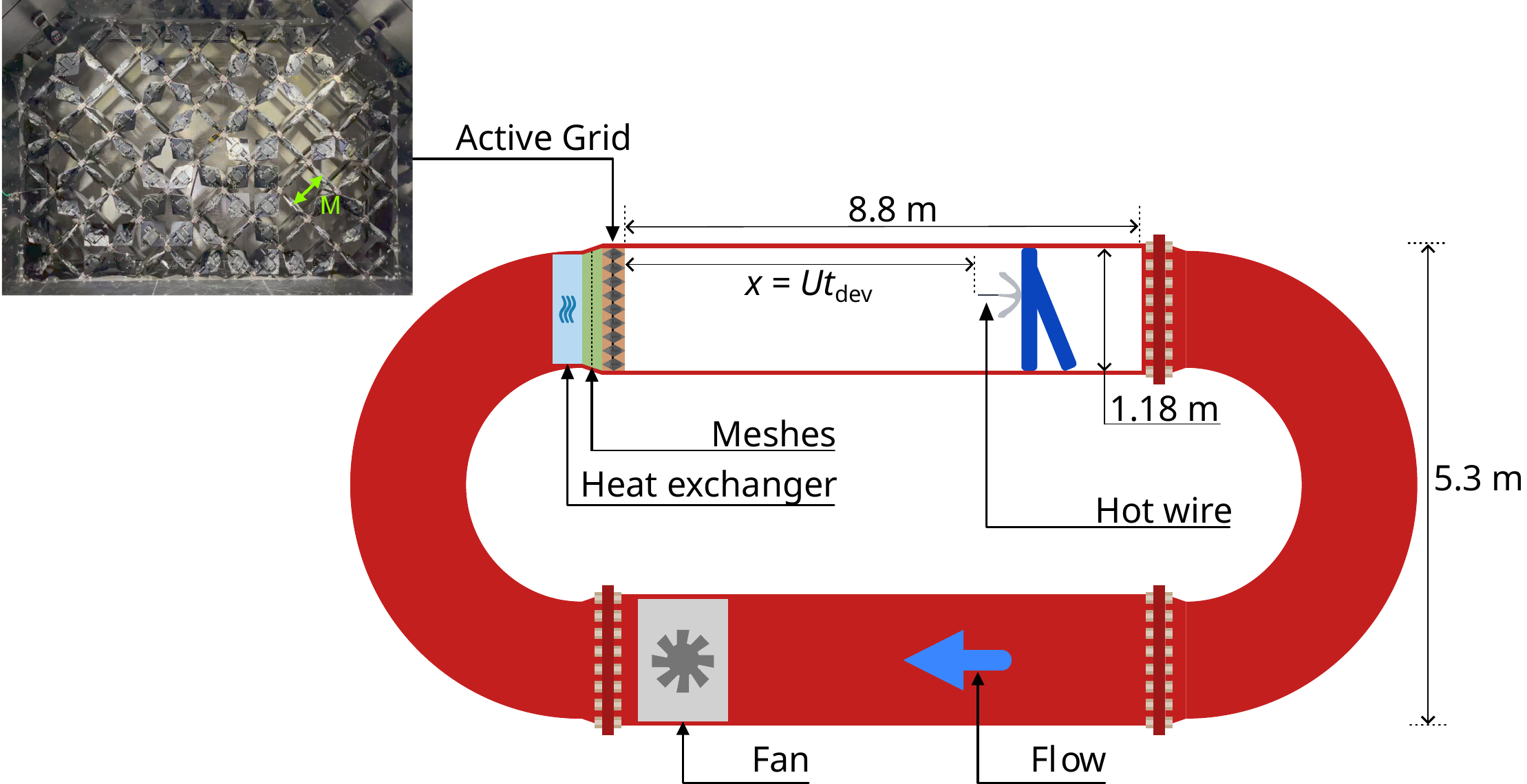}
	\caption{\textbf{Schematic of the VDTT and picture of the active grid.} \textit{Figure adapted from \cite{grunwald_effect_2026}, licensed under CC-BY 4.0.}}
	\label{fig:schematic_VDTT}
\end{figure}

\textbf{Experimental Facility—} Experiments were carried out in the Variable Density Turbulence Tunnel (VDTT) at the Max Planck Institute for Dynamics and Self-Organization, represented in Fig. \ref{fig:schematic_VDTT}. The VDTT is a closed-loop pressure vessel and wind tunnel that can be pressurised up to 19 bar. The vessel is 18.2 m long and 5.3 m high, and its total volume is 88~m$^3$. The Reynolds number of the flow Re$_\lambda$ can be changed by four different parameters: (i) the mean wind speed, $U$, controlled by the fan, which can varied from 0 to approximately 5.5 m/s, (ii) the tunnel's pressure $p$, which has direct effect on the kinematic viscosity, $\nu$ (iii) the gas. By using different gases inside the tunnel, we have control over $\nu$. Lastly, (iv) the active grid is used to increase $u^\prime$ and control the energy injection, $\epsilon$. In these experiments we varied all four parameters. See Supplementary Text and Table \ref{tab:list_exps} for reference.
In particular, turbulence is controlled by the active grid (AG) composed of 111 independently-controlled flaps managed through different grid protocols that control their spatial and temporal correlation or lack thereof. Each protocol consists of a series of typically 60 to 300 grid states, i.e. angular position of each of the flaps. Various protocols are used, some of which are extensively characterised in \cite{kuchler_measurements_2020}. The grid states are successively applied to the grid at a frequency $f_\text{grid} = 0.7$ to $10$ Hz. This makes it possible to finely tune the flow conditions in terms of streamwise and transverse correlations lengths, turbulence intensity and energy injection rate \cite{griffin_control_2019, kuchler_experimental_2019}. It should be noted that before the AG, the flow passes through a heat exchanger and a series of meshes that homogenize the flow before turbulence is produced. Small-scale eddies are expected to arise from these meshes, which are then transported through the AG down to the measurement section. The streamwise component of the velocity $v_x(t) = U + u(t)$ is recorded using a commercial hot wire anemometer (see paragraph \textit{Data acquisition}). Unless specified otherwise, the probe is located at $x = 8.5\,\mathrm{m} = 77.3M$ downstream of the active grid, where $M = 11$ cm is the size of a flap of the grid.

\noindent\textbf{Flow characteristics—}The flow is characterised by its r.m.s. velocity, $u' \equiv (\overline{u^2})^{1/2}$ (overlines indicate time averages), its energy dissipation rate $\epsilon$, its integral time scale $\tau = u'^2/\epsilon$, its Taylor scale $\lambda \equiv (u'^2/\overline{(\partial_xu)^2})^{1/2}$ and its dimensionless Reynolds number Re$_\lambda \equiv u'\lambda/\nu$, where $\nu$ is the kinematic viscosity. The turbulence intensity $u'/U$ is always less than 12\% (less than 7\% in 80\% of our experiments) and we use Taylor's hypothesis to deduce the spatial signal $u(x) \approx u(t=-x/U)$ \cite{Pope2000}. By varying the mean flow speed from $U = 1.2$ to $5.5$ m/s, the facility pressure from 1 to 6 bars, and by using either air or sulphur hexafluoride ($\mathrm{SF}_6$) we achieve Reynolds numbers ranging from $150$ to $4500$. All 36 experiments carried out are specified in the Supplementary Information. Temperature and pressure vary by less than 1\% throughout the experiments. The energy dissipation rate was calculated following results from \cite{schroder_estimating_2024}. For Re$_\lambda \lesssim 450$, where the resolution allows to compute the velocity gradients \cite{kuchler_measurements_2020}, $\epsilon = 15\nu\overline{(\partial_xu)^2}$. For higher Re$_\lambda$ where the inertial range is better defined, $\epsilon^{2/3} = C_\epsilon^{-1}\max_k [E(k)k^{5/3}]$, with $C_\epsilon = 0.49$.

\noindent\textbf{Data acquisition—}  The one-dimensional velocity field was acquired using commercial hot-wires (HW) P11 from \textit{Dantec Dynamics}, Denmark with a sampling rate of either $30$ kHz or $60$ kHz. The hot-wires were calibrated using a pitot tube connected to a differential pressure transmitter (DPT) from \textit{Siemens} before each experiment. All signals (temperature, pressure, hot-wire and DPT) were recorded using a National Instruments PCI-6123 DAQ card. The velocity signal measured from the HW is filtered by a low-pass Butterworth filter whose cut-off frequency $f_\text{cut-off}$ is the frequency above which noise dominates the energy spectrum of the unfiltered signal. The signal is then downsampled so as to reduce the sampling frequency to at most 8, but at least 4 times the cut-off frequency.

\noindent\textbf{Sampling rate correction—}In order to precisely decompose the full signal (of duration $NT$ in the case of periodic experiments) into $N$ individual realisations $u_n$, the timers which control the active grid and the DAQ card must run at the exact same pace, up to a relative precision of order 1 grid state per experiment $= f_\text{grid}^{-1}/NT \sim 10^{-6}$. However, the actual precision of these timers is not enough and one may drift with respect to the other. This mean drift is measured and corrected. To do so, the same DAQ card is used to record the input voltage of one of the flaps of the AG. This signal is periodic in the case of periodic forcing. We identify its period \textit{according to the DAQ card} and know its period \textit{according to the AG control}. The difference between the two corresponds to a small correction in (say) the sampling rate of the DAQ card. The relative discrepancy is always very small ($\sim2\times10^{-5}$) and well within the error bounds of the DAQ card specifications.

\noindent\textbf{Velocity correction—}Throughout the very long measurements (up to 50 hours), we note that the local mean of the velocity signal $v_x(t)$ varies by typically 1\%. Ref. \cite{kuchler_measurements_2020} attributed this variation to fluctuations in the HW calibration curve with time. To correct for this, the mean velocity is computed for each hour of experiment, and the ``instantaneous mean velocity $U(t)$" is obtained from this data by linear interpolation. Since $u'/U \ll 1$, the fluctuations of the HW calibration curve result in a dilatation of the velocity signal by a constant, at first order. Therefore, the signal $v_x(t)$ is corrected at each $t$ by a factor $U(0)/U(t)$. Then, the ensemble average of the first $N/2$ realisations shows no systematic discrepancy (no shift, no dilatation) with respect to the ensemble average of the last $N/2$ realisations. This confirms the relevance of this correction.

\noindent\textbf{Energy spectrum—}The energy spectra presented in the main text are not filtered nor smoothed. The energy spectrum of one experiment (see Fig. 2 of the main text) is defined by $E(f) \equiv N^{-1}\sum_{n=1}^NE_{u_n}(f)$, where the $E_{u_n}(f)$s are the power spectral density of each of the $N$ realisations of this experiment. The energy spectrum of $\langle u\rangle_N$ is directly the power spectral density of this quantity.

\clearpage

\renewcommand{\thefigure}{S\arabic{figure}}
\renewcommand{\thetable}{S\arabic{table}}
\renewcommand{\theequation}{S\arabic{equation}}
\renewcommand{\thepage}{S\arabic{page}}
\setcounter{figure}{0}
\setcounter{table}{0}
\setcounter{equation}{0}
\setcounter{page}{1} 

\section*{Supplementary Information:\\On the repeatability of turbulence} 

\subsubsection*{List of experiments}

Table \ref{tab:list_exps} lists all the experiments carried out. 7 sets of experiments have been performed:
\begin{enumerate}
    \item Experiments 1\_1 to 1\_8 investigate the influence of spatial correlations in the grid protocol;
    \item Experiments 2\_1 to 2\_5 are identical in all but the grid frequency;
    \item Experiments 3\_1 to 3\_5 correspond to the procedure described in Fig. 3 of the main text;
    \item Experiments 4\_1 to 4\_3 are analogous to 1\_1-1\_8 but use protocols from \cite{kuchler_measurements_2020}.
    \item Experiments 5\_1 and 5\_2 checked that the results are unaffected by the periodicity of the forcing. In experiment 5\_1, the $N$ realisations of duration $T$ were not done in a row, but separated by $T$. During this delay, the flow was excited by a random grid forcing sharing the same average closure and time and spatial correlations as the protocol of interest. Experiment 5\_2 is similar, except for the delays which were random. The $n$-th delay was of duration $m_nT$, where $m_n$ is an integer drawn at random with probability $p(m_n=j) = (\frac{1}{2})^j$ for $j\geq1$.
    \item Experiments 6\_1 to 6\_8 were meant to extend the range of $L$ and Re$_\lambda$ explored.
    \item Experiments 7\_1 to 7\_5 are identical in all but the probe position $x$, which varied for this dataset only. It is given in units grid mesh $M = 0.11$ m.
\end{enumerate}

Fig. \ref{fig:phaseAverage}, \ref{fig:spectrum}A, \ref{fig:spectrum}B and \ref{fig:spectrum}E display the example of experiment 1\_8. Fig. \ref{fig:spectrum}D displays the example of experiments 4\_1, 6\_5, 1\_1, 1\_7 and 1\_8. Fig. \ref{fig:ReInvariance} and \ref{fig:error_spectrum_Re} are based on dataset 3. Fig. \ref{fig:time_evolution}, \ref{fig:error_spectrum_time} and \ref{fig:time_evolution_corr} are based on dataset 7.

Values of $\epsilon$ marked with a star ($*$) are computed with the spectrum method, others are from the gradient method (see Methods). Grid protocol names are formatted as follows: spatial correlations/time correlations/r.m.s. angle of flaps/number of states. ``None" indicates no correlation. ``Gaus2/LT5/..." indicates that each flap is typically correlated to its neighbors up to two flaps away, and correlated to its previous and next 5 positions in time, with Gaussian spatial correlations and Long Tail (LT) time correlations. See \cite{kuchler_measurements_2020} for details about the grid correlations. Experiments whose number is marked with a dagger ($\dag$) have a very anisotropic grid protocol, that is, the protocol has a streamwise-to-transverse correlation lengths ratio larger than 5.

\clearpage
\newpage

\begin{table}
\hspace*{-1cm}
\rotatebox{90}{
\begin{minipage}{\textheight}

\caption{\textbf{List of the experiments.} See the Supplementary text for details. Parameters are defined in the main text and in the Methods.}
\label{tab:list_exps}

\small
\linespread{1}

\renewcommand{\arraystretch}{1.05}
\begin{tabular}{lcccccccccccccc}
\hline
\textbf{\begin{tabular}[c]{@{}c@{}}Exp.\\ \#\end{tabular}} & \textbf{Grid protocol} & \textbf{\begin{tabular}[c]{@{}c@{}}\# of\\ realisations\\ $N$\end{tabular}} & \textbf{\begin{tabular}[c]{@{}c@{}}Grid\\ freq.\\ (Hz)\end{tabular}} & \textbf{\begin{tabular}[c]{@{}c@{}}Probe\\ position\\ $x/M$\end{tabular}} & \textbf{Gas} & \textbf{\begin{tabular}[c]{@{}c@{}}Pressure\\ (bar)\end{tabular}} & \textbf{\begin{tabular}[c]{@{}c@{}}$\nu$\\ ($10^{-5}\mathrm{m^2/s}$)\end{tabular}} & \textbf{\begin{tabular}[c]{@{}c@{}}$f_\mathrm{cutoff}$\\ (Hz)\end{tabular}} & \textbf{\begin{tabular}[c]{@{}c@{}}$U$\\ (m/s)\end{tabular}} & \textbf{\begin{tabular}[c]{@{}c@{}}$u'/U$\\ (\%)\end{tabular}} & \textbf{Re$_\lambda$} & \textbf{\begin{tabular}[c]{@{}c@{}}$\epsilon$\\ ($\mathrm{m^2/s^3}$)\end{tabular}} & \textbf{$L$ (m)} & \textbf{$e_\mathrm{rand}/e_\mathrm{tot}$} \\ \hline
1\_1 & gaus2/none/rms30/100 & 1,798 & 10 & 77.3 & air & 1.01 & 1.52 & 2,054 & 4.48 & 5.00 & 258 & 0.0373 & 0.301 & 0.77 \\
1\_2 & gaus2/none/rms30/100 & 1,798 & 10 & 77.3 & air & 1.01 & 1.52 & 2,151 & 4.48 & 4.83 & 247 & 0.0355 & 0.285 & 0.77 \\
1\_3 & gaus0.54/none/rms10/100 & 1,798 & 10 & 77.3 & air & 1.02 & 1.50 & 1,907 & 4.59 & 3.23 & 151 & 0.0213 & 0.153 & 0.86 \\
1\_4 & gaus2/none/rms30/200 & 106 & 10 & 77.3 & air & 1.01 & 1.52 & 2,036 & 4.53 & 5.30 & 302 & 0.0360 & 0.384 & 0.63 \\
1\_5 & gaus2/none/rms30/250 & 998 & 10 & 77.3 & air & 0.99 & 1.55 & 1,364 & 4.50 & 4.71 & 258 & 0.0296 & 0.323 & 0.67 \\
1\_6 & gaus2/none/rms30/600 & 800 & 10 & 77.3 & air & 0.99 & 1.55 & 1,424 & 4.59 & 4.68 & 266 & 0.0293 & 0.339 & 0.66 \\
1\_7 & gaus2/none/rms30/60 & 12,500 & 10 & 77.3 & air & 1.01 & 1.52 & 1,803 & 4.46 & 4.58 & 251 & 0.0272 & 0.313 & 0.69 \\
1\_8 & gaus2/none/rms30/60 & 29,998 & 10 & 77.3 & air & 2.33 & 0.657 & 3,207 & 4.71 & 4.43 & 383 & 0.0296 & 0.307 & 0.71 \\ \hline
2\_1 & gaus2/none/rms30/150 & 200 & 10 & 77.3 & air & 1.01 & 1.52 & 1,992 & 4.38 & 4.77 & 238 & 0.0331 & 0.275 & 0.68 \\
2\_2 & $\vert$ & $\vert$ & 5 & 77.3 & air & 1.02 & 1.50 & 2,126 & 4.31 & 5.41 & 289 & 0.0355 & 0.357 & 0.57 \\
2\_3 & $\vert$ & $\vert$ & 2.5 & 77.3 & air & 1.01 & 1.52 & 2,014 & 4.37 & 5.78 & 349 & 0.0330 & 0.489 & 0.46 \\
2\_4$^\dag$ & $\vert$ & $\vert$ & 1.25 & 77.3 & air & 0.98 & 1.57 & 2,168 & 4.37 & 6.50 & 422 & 0.0348 & 0.657 & 0.38 \\
2\_5$^\dag$ & $\vert$ & $\vert$ & 0.667 & 77.3 & air & 1.01 & 1.52 & 2,195 & 4.42 & 6.91 & 483 & 0.0368 & 0.773 & 0.34 \\ \hline
3\_1 & gaus2/none/rms30/150 & 998 & 3 & 77.3 & air & 1.00 & 1.53 & 944 & 2.56 & 5.30 & 223 & 0.0067 & 0.373 & 0.56 \\
3\_2 & $\vert$ & $\vert$ & 3.87 & 77.3 & air & 0.99 & 1.55 & 1,553 & 3.46 & 5.36 & 269 & 0.0158 & 0.405 & 0.56 \\
3\_3 & $\vert$ & $\vert$ & 5 & 77.3 & air & 1.02 & 1.49 & 2,173 & 4.53 & 5.48 & 325 & 0.0361 & 0.423 & 0.57 \\
3\_4 & $\vert$ & $\vert$ & 5.88 & 77.3 & air & 1.00 & 1.53 & 2,911 & 5.37 & 5.42 & 347 & 0.0584 & 0.423 & 0.57 \\
3\_5 & $\vert$ & $\vert$ & 3.87 & 77.3 & SF6 & 5.95 & 0.0392 & 5,823 & 3.52 & 5.24 & 1685 & 0.0157* & 0.402 & 0.54 \\ \hline
4\_1 & LT3.5/LT3.5/rms50/150 & 118 & 10 & 77.3 & air & 0.99 & 1.55 & 1,971 & 3.44 & 8.09 & 311 & 0.0487 & 0.441 & 0.61 \\
4\_2 & LT3.5/LT3.5/rms50/150 & $\vert$ & 10 & 77.3 & air & 0.99 & 1.55 & 1,570 & 3.32 & 6.07 & 263 & 0.0231 & 0.355 & 0.57 \\
4\_3$^\dag$ & LT3.5/LT3.5/rms50/150 & $\vert$ & 5 & 77.3 & air & 0.99 & 1.55 & 3,351 & 5.07 & 9.47 & 509 & 0.2550* & 0.436 & 0.56 \\ \hline
5\_1 & gaus2/none/rms30/300 & 198 & 5 & 77.3 & air & 2.34 & 0.654 & 3,315 & 4.74 & 5.36 & 455 & 0.0461* & 0.355 & 0.56 \\
5\_2 & gaus2/none/rms30/100 & 200 & 5 & 77.3 & air & 2.33 & 0.657 & 3,140 & 4.74 & 5.39 & 469 & 0.0446* & 0.376 & 0.51 \\ \hline
6\_1 & gaus6/none/rms60/100 & 178 & 10 & 77.3 & air & 1.01 & 1.52 & 2,317 & 4.37 & 7.29 & 391 & 0.0661 & 0.488 & 0.59 \\
6\_2$^\dag$ & gaus1.7/none/rms30/80 & 148 & 1 & 77.3 & air & 1.01 & 1.52 & 704 & 2.12 & 6.58 & 304 & 0.0040 & 0.671 & 0.42 \\
6\_3$^\dag$ & gaus0.54/none/rms10/100 & 146 & 1 & 77.3 & SF6 & 6.00 & 0.0389 & 3,655 & 1.48 & 3.11 & 412 & 0.0010* & 0.096 & 0.72 \\
6\_4 & gaus0.4/none/rms7.5/100 & 998 & 10 & 77.3 & SF6 & 5.94 & 0.0393 & 4,007 & 2.95 & 2.65 & 621 & 0.0037* & 0.130 & 0.87 \\
6\_5 & LT7/LT5/rms50/150 & 750 & 10 & 77.3 & SF6 & 5.95 & 0.0392 & 12,429 & 4.11 & 9.89 & 2453 & 0.1740* & 0.387 & 0.62 \\
6\_6 & LT7/LT5/rms50/150 & 1,000 & 10 & 77.3 & SF6 & 5.95 & 0.0392 & 11,397 & 4.18 & 9.43 & 3137 & 0.0941* & 0.652 & 0.50 \\
6\_7$^\dag$ & LT7/LT5/rms60/150 & 998 & 5 & 77.3 & SF6 & 5.95 & 0.0392 & 14,308 & 4.20 & 10.7 & 3129 & 0.1570* & 0.571 & 0.56 \\
6\_8$^\dag$ & LT7/LT5/rms50/150 & 500 & 5 & 77.3 & SF6 & 5.94 & 0.0393 & 11,647 & 4.21 & 12.0 & 4577 & 0.1180* & 1.087 & 0.36 \\ \hline
7\_1 & gaus0.54/none/rms30/100 & 4,998 & 10 & 22.7 & air & 1.00 & 1.53 & 2,390 & 4.58 & 6.94 & 198 & 0.2550 & 0.126 & 0.78 \\
7\_2 & $\vert$ & 998 & 10 & 36.4 & air & 1.00 & 1.53 & 2,427 & 4.34 & 6.32 & 207 & 0.1304 & 0.159 & 0.77 \\
7\_3 & $\vert$ & 998 & 10 & 50.0 & air & 1.00 & 1.53 & 2,041 & 4.45 & 5.15 & 203 & 0.0660 & 0.182 & 0.79 \\
7\_4 & $\vert$ & 3,994 & 10 & 63.6 & air & 1.00 & 1.53 & 1,829 & 4.44 & 4.80 & 211 & 0.0454 & 0.214 & 0.80 \\
7\_5 & $\vert$ & 1,798 & 10 & 77.3 & air & 1.01 & 1.52 & 1,997 & 4.37 & 5.16 & 250 & 0.0412 & 0.279 & 0.76 \\ \hline
\end{tabular}
\end{minipage}
}
\end{table}

\clearpage
\newpage


\subsubsection*{Detailed model of the purely non-repeatable flow}


We recall that the same active grid forcing is reproduced $N$ times, possibly but not necessarily in a periodic manner. This results in $N$ flow realisations of duration $T$ each, and $N$ measurements $u_n(t),~n=1,\dots N$. The following calculations are carried out in the frequency space, but they should equally hold in the wavector space with the transformation $k = 2\pi f/U$, since the velocity fluctuation signal $u(t)$ can be interpreted as a spatial cut $u(x')$ where $x' = -tU$.
We have defined the ensemble-average velocity signal:

\begin{equation}
    \label{eq:phaseavg}
    \langle u \rangle_{N}(t) = \frac{1}{N}\sum_{n=1}^{N} u_n(t)
\end{equation}

We denote $E(f)$ the (one-dimensional) energy spectrum of one typical realisations of the flow, and $E_{\langle u\rangle_{N}}$ that of the ensemble-averaged velocity signal. In the following, $\langle\,\cdot\,\rangle_{N}$ indicates ensemble averaging (over a finite set of realisations), whereas $\mathbb{E}[\,\cdot\,]$ indicates the expectancy of a random variable. $\hat{g}(f)$ denotes the Fourier transform of the function $g(t)$. We assume that there exist a time scale $\tau_R \equiv 2\pi/f_R$ such that, for the smaller, faster scales $f>f_R$:

\vspace{-0.4cm}

\begin{enumerate}
    \item[(i)] \textbf{The energy spectrum of each of these realisation is the same}. This assumption is supported by the efficient model spectra which are commonly used to describe turbulence \cite{Pope2000}, but does not rely on one specific model spectrum. This implies:
    \begin{equation}
        \label{eq:hypothesis_spectrum}
        \forall f>f_R,~\forall n, ~|\hat{u}_n(f)| = \sqrt{\frac{1}{2}E(f)}
    \end{equation}
    \item[(ii)] \textbf{The phases $(\theta_n(f))_{n\in[1,N]}$ of each of the Fourier modes are independent, identically distributed (i.i.d.), uniform random variables}. In the case of independent realisations of the experiment, the independence and identical distribution of the $\theta$'s is trivial. In the case where the different realisations are all done in a row, making the entire process periodic with period $T$, it is still reasonable to assume independence and identical distribution as $T$ is much larger than any flow timescale (typically $T\geq4\tau$) and in particular than the small timescales (typically $T>10^2 f_R^{-1}$). The hypothesis of uniformity, however, is not trivial. It corresponds to the maximum randomness that can be achieved in the flow.
    \begin{equation}
        \label{eq:hypothesis_phases}
        \forall f>f_R,~\forall n,~ \theta_n \stackrel{i.i.d.}{\hookrightarrow} \mathcal{U}([0,2\pi])
    \end{equation}
\end{enumerate}
The ensemble-averaged velocity $\langle u\rangle_{N}$ reads, using hypothesis (i):
\begin{align}
     \langle u\rangle_{N} (t)& = \frac{1}{N} \sum_{n=1}^{N} u_n(t) \\
     & = \frac{1}{N} \sum_{n=1}^{N} \sum_f \hat{u}_n(f) e^{2i\pi ft} \\
     & = \sum_f \sqrt{\frac{1}{2}E(f)}\left(\frac{1}{N} \sum_{n=1}^{N} e^{i\theta_n(f)} \right) e^{2i\pi ft}
\end{align}
and we thus identify:
\begin{equation}
E_{\langle u\rangle_{N}}(f) = E(f) \left| \frac{1}{N} \sum_{n=1}^{N} e^{i\theta_n(f)} \right|^2
\end{equation}
Now, we readily obtain an expression for the attenuation factor of the energy spectrum at scales less than $L_D$:
\begin{align}
R_{N}(f) \equiv \frac{E_{\langle u\rangle_{N}}(f)}{E(f)} & = \left| \frac{1}{N} \sum_{n=1}^{N} e^{i\theta_n(f)} \right|^2 \label{eq:R}\\
& = \left(\frac{1}{N} \sum_{n=1}^{N} \cos(\theta_n(f))\right)^2 + \left(\frac{1}{N} \sum_{n=1}^{N} \sin(\theta_n(f))\right)^2
\end{align}
Let:
$$
X \equiv \frac{1}{N} \sum_{n=1}^{N} \cos(\theta_n(f)), ~~~
Y \equiv \frac{1}{N} \sum_{n=1}^{N} \sin(\theta_n(f)), ~~~
Z = \begin{pmatrix} X \\ Y \end{pmatrix}
$$
Since $\langle\cos(\theta_n)\sin(\theta_m)\rangle = \langle\cos(\theta_n)\rangle\langle\sin(\theta_m)\rangle = 0$ for $n\neq m$ and $ = \langle\frac{1}{2}\sin(2\theta_n)\rangle = 0$ for $n=m$ (assumption (ii)) $Z$ has the mean and covariance matrices:
$$
\mu = \begin{pmatrix} 0 \\ 0 \end{pmatrix}, ~~~ \sigma^2 = \begin{pmatrix} \frac{1}{2N} & 0  \\ 0 & \frac{1}{2N} \end{pmatrix}
$$
According to the central limit theorem and using assumption (ii), as $N \rightarrow +\infty$, the distribution of Z approaches the bivariate normal law $\mathcal{N}(\mu, \sigma)$. Note that, although $X$ and $Y$ are not independent (only uncorrelated), $X$ and $Y$ are asymptotically independent, since uncorrelation implies independence for a multivariate normal law. We can now rewrite:
\begin{equation}
    R_{N}(f) = X^2 + Y^2 = \frac{1}{2N}\left(\tilde{X}^2 + \tilde{Y}^2\right)
\end{equation}
where $\tilde{X} = \sqrt{2N}X$ and $\tilde{Y} = \sqrt{2N}Y$ are independent random variables following a standard normal distribution $\mathcal{N}(0,1)$. Therefore, $\tilde{R} \equiv 2NR$ follows a $\chi_2^2$ (chi-squared) distribution:
\begin{equation}
    \mathbb{E}[\tilde{R}] = 2, ~~~~
    \mathrm{Var}(\tilde{R}) = 4, ~~~~
    p({\tilde{R}}) = \frac{1}{2} e^{-\tilde{R}/2}
\end{equation}
and finally, $\forall k$
\begin{equation}
    \label{eq:model-predict}
    \boxed{\mathbb{E}[R_{N}] = \frac{1}{N}} ~~~ \boxed{\mathrm{Var}(R_{N}) = \frac{1}{N^2}} ~~~ \boxed{p(R_{N}) = N e^{-NR_{N}}}
\end{equation}


\clearpage
\setcounter{figure}{0}
\section*{Extended Data:\\On the repeatability of turbulence} 

\subsubsection*{Convergence of experimental observations for $N\to\infty$}

\begin{figure}[h]
    \centering
    \includegraphics[width=\linewidth]{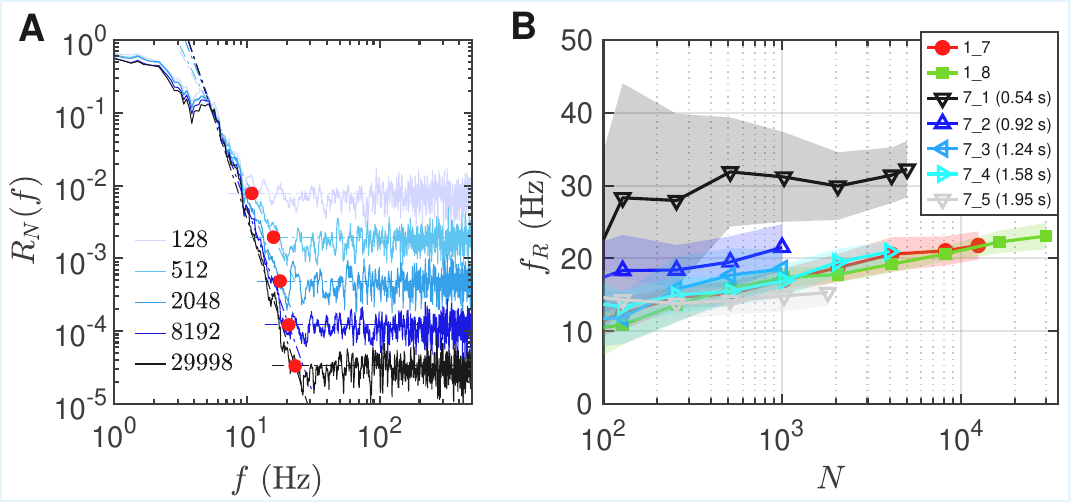}
    \caption{\textbf{Estimate of the upper random scale $f_R$}. \textbf{(A)} In one experiment (1\_8), $R_N(f)$ for several $N$s (indicated in legend), power law and constant fits (dashed lines) and their intersection defining the best estimate of $f_R$ for a given sample size $N$ (red dots). \textbf{(B)} Estimate of $f_R$ for the 2 longest experiments and the streamwise profile. The legend indicates the experiment number (as in Table \ref{tab:list_exps}) and, for the streamwise profile, the flow development time. Shaded areas indicate the error bars (see Supplementary Text for a definition).}
    \label{fig:f_R}
\end{figure}

Here we discuss the convergence of the proposed statistics $f_R$ and $e_\text{rand}/e_\text{tot}$ as $N\to\infty$.
In order to calculate $f_R$, we fit $R_N(f)$ by a power law in the range of $f$'s such that $R_N(0)/3 > R_N(f) > 3/N$. $f_R$ for this $N$ is estimated by the x-coordinate of the intersection of the latter curve with $1/N$. Fig. \ref{fig:f_R} presents $f_R(N)$ for 7 experiments. Error bars correspond to the standard error on the fit parameters plus the observed up-to-5\% deviation of $\langle {R_N(f>f_R)}\rangle$ from $1/N$. The larger the $N$, the more precise the ensemble statistics are, and the more repeatable components can be identified, even when their predictability is very weak. Therefore, $f_R(N)$ increases with $N$ and seems to converge. This value is referred to as $f_R$ in the main text. Additionally, one can see in Fig. \ref{fig:f_R} that very large ensemble size, at least $N \gtrsim 10^4$, are necessary to estimate the large-$N$ limit. This size is only achieved in two experiments (1\_7 and 1\_8). In the absence of such long statistics for 2 experiments of dataset 7, and for the sake of consistency Fig. \ref{fig:time_evolution} of the main text shows $f_R(N=10^3)$. $f_R(10^3)$ is quantitatively not equal to the $N\to\infty$ limit (see for example curve 7\_2), but has sufficiently small error bars to assess the \textit{trend} of the evolution of $f_R$ with development time.

\begin{figure}[h!]
    \centering
    \includegraphics[width=0.45\linewidth]{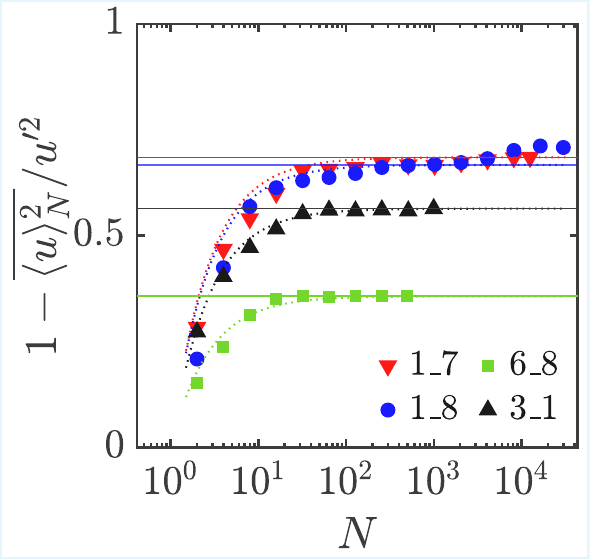}
    \caption{\textbf{Estimate of $e_\mathrm{rand}/e_\mathrm{tot}$ as a function of $N$ for four experiments} (see Table \ref{tab:list_exps}). Horizontal lines indicate the estimate of the limit. For reference, dotted line indicates the curve $e_\mathrm{rand}/e_\mathrm{tot}(1-N^{-1})$.} 
    \label{fig:ErandEtot_convergence}
\end{figure}

Fig. \ref{fig:ErandEtot_convergence} displays the quantity $1-\overline{\langle u\rangle_N^2}/u'^2$ as a function of $N$, for various experiments (different $Re_\lambda$, grid protocol...). By definition, it tends to $e_\mathrm{rand}/e_\mathrm{tot}$ as $N\to\infty$. Fig. \ref{fig:ErandEtot_convergence} shows that this limit is approached to a few percentage points for $N\gtrsim 100$.
If the components at each frequency $f$ were either entirely random, either entirely repeatable, i.e. $\forall N,~\forall f<f_R(N),~\mathbb{E}[R_N(f)] = 1$ (and there were no range where $R_N(f)$ behaves like a power law of $f$), then:
\begin{enumerate}
    \item[(i)] $e_\mathrm{tot} = \frac{1}{2}u'^2$
    \item[(ii)] $e_\mathrm{rand} = \sum_{f>f_R} E(f)$
    \item[(iii)] $\frac{1}{2}\overline{\langle u\rangle_N^2} \approx \sum_{f<f_R} E(f) + \frac{1}{N}\sum_{f>f_R}E(f) = (e_\mathrm{tot}-e_\mathrm{rand}) + \frac{1}{N}e_\mathrm{rand}$, where in the last sum the attenuation of each random component has been approximated by the average attenuation $1/N$.
\end{enumerate} 
In this simplified picture, $1-\overline{\langle u\rangle_N^2}/u'^2 = (1-1/N)e_\mathrm{rand}/e_\mathrm{tot}$. For reference, this behaviour is depicted by the dotted curves.
In one case (1\_8) $1-\overline{\langle u\rangle_N^2}/u'^2$ unexpectedly varies significantly at $N \approx 10^4\gg 100$. This is most likely due to changes in the experiment, for Experiment 1\_8 is by far the longest of the present work (50 hours) and it is quite possible that the hot wire calibration, the pressure or the temperature varies significantly at some point.

\newpage
\subsubsection*{Connection between $e_\text{rand}/e_\text{tot}$ and the error energy}

Many theoretical and numerical studies of the predictability of turbulence introduce the error energy spectrum \cite{boffetta_chaos_2017, berera_chaotic_2018, lorenz_predictability_1969, Lorenz1963, leith_predictability_1972, ge_production_2023, metais_statistical_1986}, here defined as:
\begin{equation}
    \Delta(f) = \langle |\widehat{\delta u}_{n,m}|^2\rangle
\end{equation}
where $\delta u_{n,m}(t) = (u_n(t) - u_m(t))/\sqrt{2}$ and the average is taken over all pairs. Note that in these studies $\Delta$ is actually defined in the wavenumber space $k$, but in the present study this is equivalent to our frequency space $f$ under Taylor's frozen flow hypothesis. 

The ensemble-averaging approach presented here and this error energy spectrum approach are closely connected (as remarked in Ref. \cite{vela-martin_predictability_2024}). The energy measured by $\Delta$ is that of the components that are found in one or some realisations but not all, and these components are precisely removed (or partially removed) when averaging over all realisations. In particular, we show here that the total error energy $e_\Delta \equiv \int\mathrm{d}f\Delta(f)$ is equal to $e_\text{rand}$ up to a vanishing error:
\begin{equation}
    e_\Delta - e_\text{rand} = \frac{1}{N}\left(e_\text{tot} - \mathbb{E}[C_{n, m\neq n}(0)] \right) + O\left(\frac{1}{N^2}\right) = O\left(\frac{1}{N}\right)
    \label{eq:Erand-Edelta}
\end{equation}
where $C_{n,m}(\Delta t)$ is the cross-correlation of $u_n$ and $u_m$. Indeed:
\begin{align}
    e_\Delta - e_\mathrm{rand} &= e_\Delta - e_\mathrm{tot} + e_{\langle u\rangle} \\
    &= \int_{-\infty}^\infty\mathrm{d}f\left[\Delta(f) - E_u(f) + E_{\langle u\rangle}(f)\right] \\
    &= \int\mathrm{d}f\left[\frac{1}{N(N-1)}\sum_{n\neq m}\frac{1}{2}|\widehat{u}_n-\widehat{u}_m|^2  - \frac{1}{N}\sum_n |\widehat{u}_n|^2 + \left|\frac{1}{N}\sum_n \widehat{u}_n\right|^2 \right] \\
    &= \int\mathrm{d}f\left[\frac{1}{N(N-1)}\sum_{n\neq m}\frac{1}{2}\left( |\widehat{u}_n|^2 +|\widehat{u}_m|^2 - 2\mathrm{Re}\{\widehat{u}_n\widehat{u}_m^*\}\right)  - \frac{1}{N}\sum_n |\widehat{u}_n|^2 + \frac{1}{N^2}\sum_{n,m} \widehat{u}_n\widehat{u}_m^* \right] \\
    &= \int\mathrm{d}f\left[-\frac{1}{N(N-1)}\sum_{n\neq m}\mathrm{Re}\{\widehat{u}_n\widehat{u}_m^*\} + \frac{1}{N^2}\sum_{n,m} \widehat{u}_n\widehat{u}_m^* \right] \\
    &= \int \frac{\mathrm{d}f}{N(N-1)}\sum_{n\neq m}\mathrm{Im}\{\widehat{u}_n\widehat{u}_m^*\} + \int\mathrm{d}f\left[ \frac{1}{N^2}\sum_n |\widehat{u}_n|^2 + \left( \frac{1}{N^2}-\frac{1}{N(N-1)} \right)\sum_{n\neq m} \widehat{u}_n\widehat{u}_m^* \right] \\
    &= \frac{1}{N(N-1)}\sum_{n\neq m}\int\mathrm{d}f\,\mathrm{Im}\{\widehat{u}_n\widehat{u}_m^*\} + \frac{1}{N}\left( e_\mathrm{tot} - \left\langle\int\mathrm{d}f\,\widehat{u}_n\widehat{u}_m^*\right\rangle_{n\neq m}\right)
\end{align}
where normalizing prefactors in the Fourier transforms have been omitted. But $\widehat{u}_n\widehat{u}_m^*$ is the Fourier transform of the cross-correlation of $u_n$ and $u_m$, $C_{n,m}(\Delta t)$. It is a real function, the imaginary part of its Fourier transform is therefore odd in $f$ and vanishes when integrated over all $f$s. The first term is thus zero. In the second term, the integral of the Fourier transform of the cross-correlation over all $f$s corresponds to the inverse Fourier-transform for $\Delta t=0$. From the central limit theorem, its average over all pairs $n\neq m$ is equal to a constant (the expectation, $<\infty$ from physical considerations), up to corrections in $O(1/N)$. This achieves the proof of Eq. \ref{eq:Erand-Edelta}.

\begin{figure}
    \centering
    \includegraphics[width=0.55\linewidth]{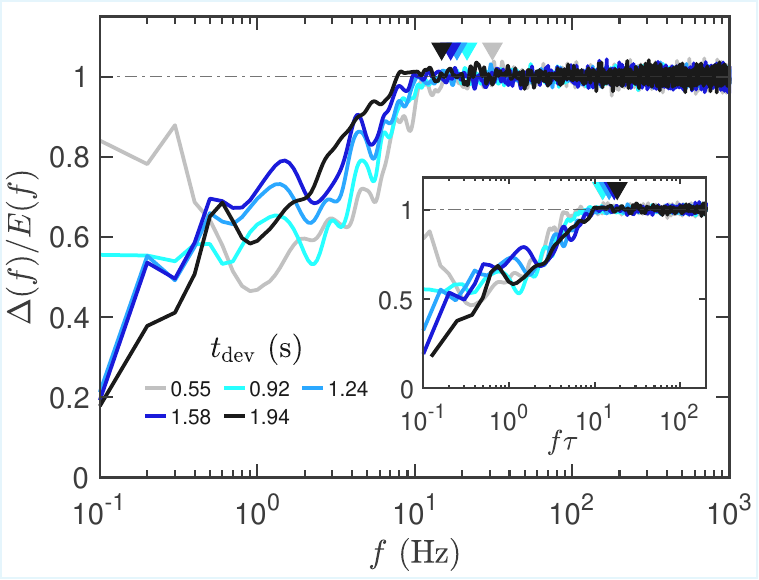}
    \caption{\textbf{Evolution of the error energy spectrum with development time.} Ratio of the error energy spectrum to the energy spectrum, for several development times, as a function of frequency. The triangles indicate $f_R$. Inset: Same but as a function of the frequency rescaled by the integral time scale.}
    \label{fig:error_spectrum_time}
\end{figure}

In Figure \ref{fig:error_spectrum_time}, we plot the ratio of the error energy spectrum to the energy spectrum $\Delta(f)/E(f)$ for several development times (smoothed by convolution with a Gaussian Kernel of 21 points). To calculate $\Delta(f)$, the mean is taken over 500 random pairs, but we checked that this was sufficient for $\Delta$ to converge. This quantity is analogous to $R_N(f)$ in the ensemble-average approach. At $f>f_R$ (small scales), it is around 1, indicating that flow realisations are completely uncorrelated. On the contrary, at large scales $f\ll f_R$, $\Delta(f)/E(f)$ is only around 0.2-0.8, indicating that at these scales the flow is predominantly repeatable. With increasing development time, the curves shift to the left as the uncertainty contaminates larger scale. However, and contrary to the case of stationary turbulence, $E(f)$ changes as well and energy becomes confined to larger scales too. The inset shows that in flow units $f\tau$, the curves do not shift to the left any more. In these units, the uncertainty does not catch the largest energy-containing scales.

\begin{figure}
    \centering
    \includegraphics[width=0.55\linewidth]{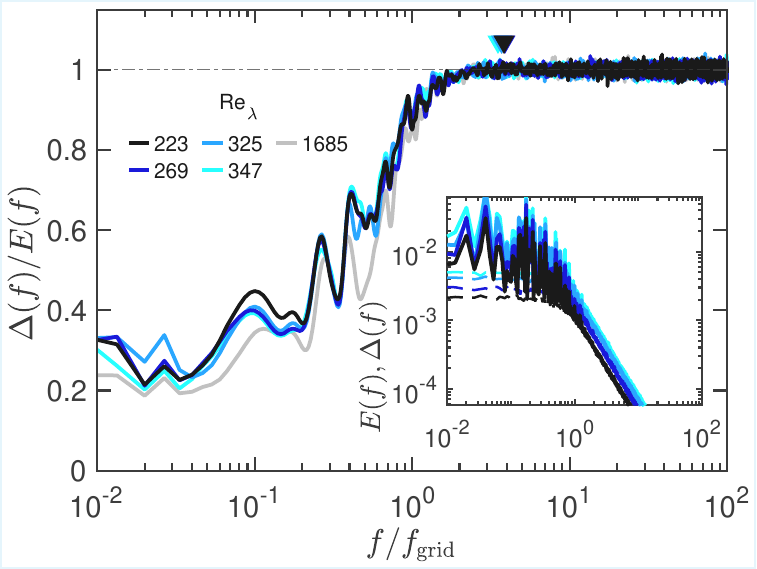}
    \caption{\textbf{Error energy spectrum and Reynolds number.} Ratio of the error energy spectrum to the energy spectrum for the same protocol but several Reynolds numbers (procedure described in the main text and in Fig. \ref{fig:ReInvariance}). In units $f/f_\text{grid}$, the five experiments have the same spatial scales, time scales and development time. The triangles indicate $f_R/f_\text{grid}$. Inset: Error energy spectra (dashed lines) and energy spectra (solid lines) as a function of $f/f_\text{grid}$.}
    \label{fig:error_spectrum_Re}
\end{figure}

Figure \ref{fig:error_spectrum_Re} shows the same quantity for the experiments of Fig. \ref{fig:ReInvariance} of the main text, which differ only by the Reynolds number provided that they are looked at in time units $f_\text{grid}^{-1}$. The curves do not shift to the left but overlap (in this rescaled units), in spite of the changing Reynolds number by a factor of 8.

\newpage

\subsubsection*{Time evolution of the ensemble-averaged signal, and absence of correlation with the grid solidity}

\begin{figure}[h]
    \centering
    \includegraphics[width=\linewidth]{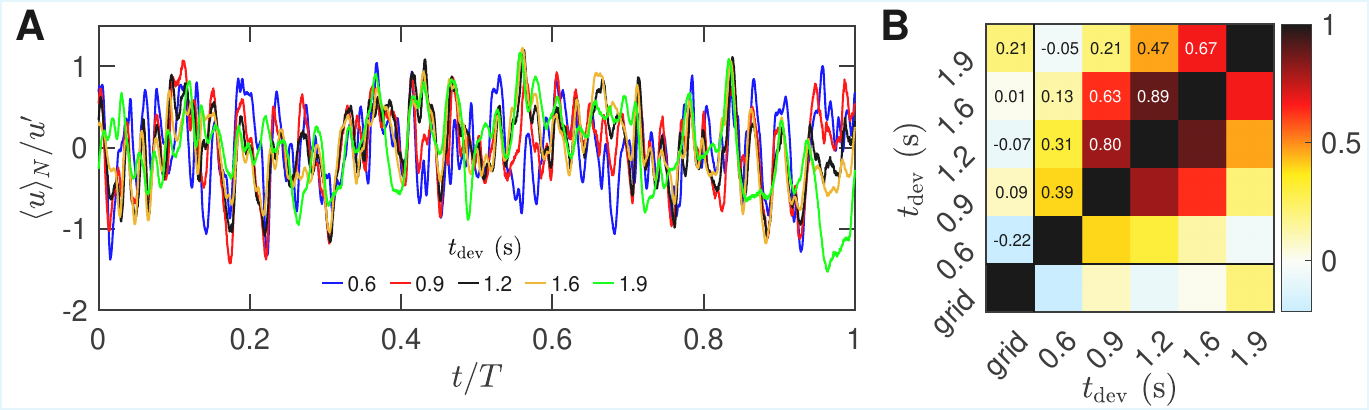}
    \caption{\textbf{Evolution of the ensemble-averaged velocity fluctuations with time.} \textbf{(A)} Ensemble-averaged velocity fluctuations normalised by $u'$ for several development times (indicated in plot), as a function of time. Time origins are shifted by the development time. \textbf{(B)} Pearson correlation coefficient between each pair of signals, and with the grid spatially-averaged aperture (defined in the Supplementary Text) at the same time.}
    \label{fig:time_evolution_corr}
\end{figure}

The observed non-zero ensemble-averaged velocity fluctuations do not correspond to a change in the mean flow of the tunnel, due to the state of the grid. Instead, they reflect more complex, spatially and temporally varying repeatable structures. This can be checked in Fig. \ref{fig:time_evolution_corr}A, where the ensemble-averaged velocity fluctuations of the same flow are plotted for different development time. The signals are complex and do not overlap completely each other, although correlations are present and suggest that large-scale coherent structures exist. Quantitatively, \ref{fig:time_evolution_corr}B shows that the ensemble-averaged velocity fluctuations measured at a given development time are correlated with those measured a bit later or earlier, but these correlations (and the associated large-scale coherent structures) only survive a few tenths of a second. On the contrary, the constant share of random energy $e_\text{rand}/e_\text{tot}$ is found to be constant on much longer time scales (at least 1.4 seconds). As a further check, we assess whether the ensemble-averaged velocity fluctuations are correlated with the average aperture of the grid, defined as $1-\langle|\sin(\varphi(t)_m)|\rangle_m$, where the $\varphi_{m=1,\dots,111}$ are the angles of each of the flaps and the average $\langle \rangle$ is taken over all flaps. This aperture ranges from 0 (all flaps closed) to 1 (all flaps fully open). If the observed non-zero ensemble-averaged velocity fluctuations where the signature of a time-varying mean flow in the whole tunnel, one would expect them to be correlated to the grid aperture at the same time. Instead, Fig. \ref{fig:time_evolution_corr} shows that the correlation is small if any, and can be equally positive or negative. It is much smaller than the correlations between signals with comparable $t_\text{dev}$ too, indicating that the latter correlations are due to large-scale coherent structures, not to mean flow. Finally, let us highlight that when looking at the cross-correlation between the grid aperture and the ensemble-average velocity fluctuations (shifted or not by $t_\text{dev}$), the zero-lag value is typically not the largest. Many other peaks of comparable size, both negative and positive, are observed.


\end{document}